\documentstyle[11pt,a4wide,amsmath,amssymb,epsf]{article}

\newcommand{\be}{\begin{equation}}
\newcommand{\ee}{\end{equation}}
\newcommand{\bea}{\begin{eqnarray}}
\newcommand{\eea}{\end{eqnarray}}

\newcommand{\nn}{\nonumber\\}

\begin{document}

\begin{center}
{\large {\bf  Can strings live in four dimensions? }} \\
\vspace{1cm}
{\bf Jean Alexandre} and {\bf Nikolaos E. Mavromatos} \\
\vspace{0.3cm}
Department of Physics, King's College London, WC2R 2LS, UK \\
\vspace{2cm}
{\bf ABSTRACT}:\\
 \vspace{0.5cm}
\end{center}
Using a novel, non-perturbative, time-dependent string configuration 
derived in [1], we present here an argument which selects new critical dimensions for the target space time
of a bosonic sigma model,
with $D=4$ the lowest non trivial value. 
This argument is based on the properties of the partition function after a target space Wick rotation.

\vspace{2cm}

In \cite{AEM}, the authors were interested in a time-dependent configuration of the bosonic string,
relevant to the description of a spatially-flat Robertson Walker Universe, with metric
$ds^2=dt^2-a^2(t)(d\vec x)^2$, where $t$ is the time in the Einstein frame, and $a(t)$ is the scale factor.  
It was argued that the following time-dependent configuration 
\be\label{config}
S=\frac{1}{4\pi\alpha^{'}}\int d^2\xi\sqrt{\gamma}\left\{\gamma^{ab}\frac{A\eta_{\mu\nu}}{(X^0)^2}
\partial_a X^\mu\partial_b X^\nu+\alpha^{'}R^{(2)}\phi_0\ln(X^0)\right\},
\ee
where $A$ and $\phi_0$ are constants, is a fixed point of the $\alpha^{'}$ flow equations derived in \cite{AEM}.
It was then conjectured that this configuration satisfies Weyl invariance conditions to all orders in 
the Regge slope $\alpha^{'}$. This result holds 
for any target space dimension $D$, and the constant $A$ depends on $\phi_0$ and $D$, in a way
that could not be determined, since this would require knowledge of 
the Weyl anomaly coefficients to all orders in $\alpha^{'}$.\\
The corresponding scale factor was then shown to be a power law
\be\label{scale}
a(t)\propto t^{1+\frac{D-2}{2\phi_0}},
\ee
such that, if the following relation holds
\be\label{Minkowski}
D=2-2\phi_0,
\ee
the target space is static and flat (Minkowski Universe).

\vspace{0.5cm}

In this note, we wish to study the properties of the configuration (\ref{config}) under the
target space Wick rotation, as appropriate for a well-defined world-sheet path integral, since in such a case
the two-dimensional field $X^0(\xi)$ does not have negative norm:
\be\label{transfo}
X^0\to iX^0.
\ee
For a given world sheet Euclidean metric, the partition function obtained from the action (\ref{config}) is
\be\label{Z}
{\cal Z}=\frac{\int{\cal D}[X^\mu]\exp(-S)}{\int{\cal D}[X^\mu]},
\ee
where $\int{\cal D}[X^\mu]=V$ is the target space volume.
The analytic continuation (\ref{transfo}) implies the expected change of the target space Minkowski metric
into a Euclidean one, since
\bea
\frac{\eta_{\mu\nu}}{(X^0)^2}\partial_a X^\mu\partial_b X^\nu
&\to&\frac{\partial_a (iX^0)\partial_b (iX^0)}{(iX^0)^2}
-\delta_{jk}\frac{\partial_a X^j\partial_b X^k}{(iX^0)^2}\nn
&=&\frac{\delta_{\mu\nu}}{(X^0)^2}\partial_a X^\mu\partial_b X^\nu.
\eea
We are therefore led to the following action
\be
S\to\tilde S=S_E+i\frac{\pi}{2}\phi_0\chi,
\ee
where the (target space and world sheet) Euclidean action is 
\be
S_E=\frac{1}{4\pi\alpha^{'}}\int d^2\xi\sqrt{\gamma}\left\{\gamma^{ab}\frac{A\delta_{\mu\nu}}{(X^0)^2}
\partial_a X^\mu\partial_b X^\nu+\alpha^{'}R^{(2)}\phi_0\ln(X^0)\right\},
\ee 
and
\be
\chi=\frac{1}{4\pi}\int d^2\xi\sqrt{\gamma}\gamma^{ab}R^{(2)}
\ee
is the Euler characteristic of the world sheet. For a closed world sheet, without cross-caps, we have \cite{GSW}
\be
\chi=2-2g,
\ee
where $g$ is the number of handles, such that the partition function becomes, after the Wick rotation
(\ref{transfo}),
\be
{\cal Z}\to\tilde{\cal Z}={\cal Z}_E\exp\{i\pi(1-g)\phi_0\},
\ee
where ${\cal Z}_E$ is the partition function corresponding to the action $S_E$.
For $\tilde{\cal Z}$ to be real, we need the quantization condition
\be
(1-g)\phi_0=n,
\ee
where $n$ is an integer, different from 0, since $\phi_0\ne 0$ in the solution of \cite{AEM}.
Together with the relation (\ref{Minkowski}), this quantization of the 
dilaton amplitude yields for the dimension $D$
\be\label{quant}
D=2-\frac{2n}{1-g},
\ee
in order to have a Minkowski target space.
If we consider the case of the spherical world sheet, with $g=0$, and for which the configuration (\ref{config}) 
was derived, the allowed dimensions are then obtained for $n<0$ and are the even integers not smaller than 4
\be\label{dimensions}
D=4,6,8,10,...,26,...
\ee
As a consequence, $D=4$ is the minimum dimension for which the analytic continuation (\ref{transfo}) 
can be performed consistently in the partition function (\ref{Z}), and for which the string propagates
in a Minkowski space time background.\\
Before closing we would like to stress the difference of our approach to that of \cite{ABEN,ddk},  as far as 
the target-space dimensionality is concerned. In those works, the target time $X^0$ \cite{ABEN} and the Liouville
mode $\varphi$ \cite{ddk} have a similar r\^ole, namely to restore the central charge of the sigma model 
theory to its critical value, fixed by the ghost fields.
This was due to the fact that the corresponding field theories in question were simple ones, with 
canonically normalized kinetic terms for the sigma model fields, where the resulting target space dimensionality
depends on the value of the dilaton and/or Liouville amplitude.\\ 
Instead, in our approach \cite{AEM}, the fields $X^\mu$ are not canonically normalized, and the
above-mentioned results do not apply. As argued in \cite{AEM}, the conformal invariance is
satisfied for any target-space dimensionality $D$, independently of the dilaton amplitude, 
provided the target space has a Minkowski signature.
However, as we have discussed here, the condition (\ref{Minkowski}) for a static target space and the 
analytic continuation (\ref{transfo}) to a Euclidean Universe imply the quantization (\ref{quant}) of $D$.\\
Our result is also different from the strongly-coupled Liouville string regime discussed in \cite{gervais}, 
for central charges $1<c<25$, where different critical values for $D=7,13,19$ are obtained. \\
We should stress once more that our approach is only based on sigma-model quantization methods.
It remains to be seen whether a consistent (super)string theory ghost-free spectrum can be 
built on the entire set of the allowed dimensions (\ref{dimensions}), stemming from our
procedure.

\end{document}